\documentclass[sigconf,screen]{acmart}

\usepackage{url}
\usepackage{xspace}
\usepackage{amsthm}
\usepackage{diagbox}
\usepackage{hyperref}
\usepackage{makecell}
\usepackage{multirow}
\usepackage{graphicx}
\usepackage{booktabs}
\usepackage{enumerate}
\usepackage{todonotes}
\usepackage{algorithm}
\usepackage{color,xcolor}
\usepackage{algpseudocode}
\usepackage{amsmath,amsfonts}

\newcommand{\ie}{\textit{i.e.,}\xspace}
\newcommand{\eg}{\textit{e.g.,}\xspace}
\newcommand{\etc}{\textit{etc.}\xspace}
\newcommand{\etal}{\textit{et al.}\xspace}

\newcommand{\figref}[1]{Fig.~\ref{#1}\xspace}
\newcommand{\tabref}[1]{TABLE~\ref{#1}\xspace}

\newcommand{\formref}[1]{Formula~\ref{#1}\xspace}

\newcommand{\monkey}{Monkey\xspace}
\newcommand{\vtest}{VTest\xspace}
\newcommand{\toolname}{{\sc RoboTest}\xspace}
\newcommand{\toolrandom}{{\sc RoboTest-Random}\xspace}
\newcommand{\tooledge}{{\sc RoboTest-Edge}\xspace}
\newcommand{\toolcenter}{{\sc RoboTest-Center}\xspace}

\begin{document}

\title{Practical Non-Intrusive GUI Exploration Testing \\ with Visual-based Robotic Arms}

\acmConference[ICSE 2024]{46th International Conference on Software Engineering}{April 2024}{Lisbon, Portugal}

\author{Shengcheng Yu}
\affiliation{\institution{State Key Laboratory for Novel Software Technology} 
\city{Nanjing University}
\country{China}}
\email{yusc@smail.nju.edu.cn}

\author{Chunrong Fang}
\authornote{Chunrong Fang is the corresponding author.}
\affiliation{\institution{State Key Laboratory for Novel Software Technology} 
\city{Nanjing University}
\country{China}}
\email{fangchunrong@nju.edu.cn}

\author{Mingzhe Du}
\affiliation{\institution{State Key Laboratory for Novel Software Technology} 
\city{Nanjing University}
\country{China}}
\email{nandodu@smail.nju.edu.cn}

\author{Yuchen Ling}
\affiliation{\institution{State Key Laboratory for Novel Software Technology} 
\city{Nanjing University}
\country{China}}
\email{yuchen_ling@nju.edu.cn}

\author{Zhenyu Chen}
\affiliation{\institution{State Key Laboratory for Novel Software Technology} 
\city{Nanjing University}
\country{China}}
\email{zychen@nju.edu.cn}

\author{Zhendong Su}
\affiliation{\institution{ETH Zurich} \country{Switzerland}}
\email{zhendong.su@inf.ethz.ch}

\renewcommand{\shortauthors}{Shengcheng and Chunrong et al.}

\begin{abstract}

Graphical User Interface (GUI) testing has been a significant topic in the software engineering community. Most existing GUI testing frameworks are intrusive and can only support some specific platforms, which are quite limited. With the development of distinct scenarios, diverse embedded systems or customized operating systems on different devices do not support existing intrusive GUI testing frameworks. Some approaches adopt robotic arms to replace the interface invoking of mobile apps under test and use computer vision technologies to identify GUI elements. However, Some challenges remain unsolved with such approaches. First, existing approaches assume that GUI screens are fixed so that they cannot be adapted to diverse systems with different screen conditions. Second, existing approaches use XY-plane robotic arm system, which cannot flexibly simulate human testing operations. Third, existing approaches ignore the compatibility bugs of apps and only focus on the crash bugs. To sum up, a more practical approach is required for the non-intrusive scenario.

In order to solve the remaining challenges, we propose a practical non-intrusive GUI testing framework with visual-based robotic arms, namely \toolname. \toolname integrates a set of novel GUI screen and widget detection algorithm that is adaptive to detecting screens of different sizes and then to extracting GUI widgets from the detected screens. Then, a complete set of widely-used testing operations are applied with a 4-DOF robotic arm, which can more effectively and flexibly simulate human testing operations. During the app exploration, \toolname integrates the specially designed Principle of Proximity-guided (PoP-guided) exploration strategy, which chooses close widgets of the previous operation targets to reduce the robotic arm movement overhead and improve exploration efficiency. Moreover, \toolname can effectively detect some compatibility bugs beyond crash bugs with a GUI comparison on different devices of the same test operations. We evaluate \toolname with 20 real-world mobile apps, together with a case study on a representative industrial embedded system. The results show that \toolname can effectively, efficiently, and generally explore the AUT to find bugs and reduce app exploration time overhead from the robotic arm movement.

\end{abstract}

%
%

\maketitle

\section{Introduction}

Graphical User Interface (GUI) testing is a significant part of guaranteeing application (app) quality \cite{said2020gui}. Automated GUI testing is one of the most effective approaches to the GUI-based app quality guarantee. Most existing approaches \cite{kong2018automated} are the imitation of human interactions with the apps under test (AUT), which consists of two parts, the target (\ie GUI widget) identification and the target operation. For target identification, existing approaches adopt tools that can obtain the GUI layout structure information during app runtime. For target operation, most approaches adopt intrusive technologies, directly operating targets via interface invoking.

Intrusive approaches are faced with several challenges. First, during the target identification, intrusive approaches may ignore some GUI widgets, like those contained in \texttt{canvas} or embedded H5 pages (\ie embedded HTML pages into a GUI container), and their inner components are invisible in the GUI layout structure \cite{yu2019crowdsourced} but are important in the apps. Second, besides Android and iOS as subjects of most GUI testing studies, apps are running on more distinct platforms, including industrial embedded systems. In order to conduct GUI testing on different environments, testing approaches should get rid of the dependency on specific operating system support, and the intrusive target identification and operation are no more usable. Moreover, some specific systems, like the embedded systems or the lightweight industrial systems, do not support existing intrusive frameworks, \eg Appium \cite{appium}. Third, some app compatibility bugs triggered by emerging diverse devices (like the motivating example in \figref{fig:screen}) are hard to detect by existing intrusive approaches since such compatibility bugs cannot be found from the software aspect.

Under such a circumstance, it is necessary to effectively simulate human interactions \cite{qian2020roscript, ran2022automated} with the combination of computer vision (CV) technologies for target identification and the robotic arm for target operation. RoScript \cite{qian2020roscript} is a state-of-the-art approach to record and replay GUI test scripts based on the robotic arm. identifying GUI widgets and the finger movements with CV technologies to record the operations and the corresponding widgets. Then the robotic arm is used to conduct the recorded operations via image matching. However, the script development still needs huge human labor, and extra manual attention is required during the recording process. Further, Ran \etal \cite{ran2022automated} propose \vtest, the state-of-the-art tool that utilizes robotic arms to conduct automated non-intrusive GUI exploration. However, \vtest still fails in some challenges. First, it does not consider the diversity of GUI screens, including mobile devices and industrial embedded systems, leading to the \vtest not being generally usable. Second, it cannot effectively simulate human testing operations with a limited XY-plane robotic arm and brings much extra overhead, greatly lowering the GUI exploration efficiency. Third, it only focuses on crash bugs and ignores the bugs triggered by the compatibility between apps and devices. 

Faced with such challenges, we present a practical non-intrusive GUI testing approach with visual-based robotic arms, \toolname. \toolname adopts a novel GUI screen detection algorithm, solving the image distortion and deflection problems with a two-step correction. From the recognized GUI screens, \toolname extracts the widgets with adjusted CV algorithms. Then, based on the identified widgets, \toolname manipulates the robotic arm to operate on the target widgets based on a set of GUI operation instructions for the robotic arm to implement widely-used GUI testing operations. In order to improve the GUI exploration efficiency, \toolname adopts an optimized Principle of Proximity-guided (PoP-guided) exploration strategy, which chooses close widgets of the previous operation targets.

GUI screens of different devices have different sizes \cite{wei2016taming}. For existing non-intrusive approaches \cite{ran2022automated, qian2020roscript}, manual efforts of the device placement and extra manual configuration are required to make sure that the robotic arm can capture the GUI screens with the camera. \toolname utilizes a novel GUI screen detection algorithm, adapted to detect devices screens of different sizes. The GUI screen detection algorithm solves the image distortion and deflection problems with a two-step correction. Then \toolname extracts GUI widgets from the app GUI, and a combination of traditional CV algorithms and deep learning (DL) models is adopted to identify the GUI widgets. Different from the widget identification algorithms \cite{yu2021layout} for intrusive approaches, which take high-resolution app screenshots as input, the GUI images taken by robotic arm cameras are less clear. Therefore, specific processing and parameter settings are adjusted to make the visual algorithm more suitable.

To implement a complete set of widely-used testing operations, we adopt a four-degrees-of-freedom (4-DOF) robotic arm and design two kinds of movements, atomic movements, and compound movements. Compared with the XY-plane robotic arm, the 4-DOF robotic arm is more flexible and can simulate human operations more realistically and efficiently. Atomic movements refer to single-direction movements, and different from the XY-plane robotic arms \cite{ran2022automated, qian2020roscript}, the operations on the 4-DOF robotic arm need to be decomposed to different degrees. We design six atomic movements in different directions: forward and backward, up and down, left and right. Compound movements are composed of atomic movements, including \texttt{click()}, \texttt{long\_click()}, \texttt{double\_click()}, \texttt{slide()}, \texttt{scroll()}, and \texttt{input()}.

During the exploration, we first apply a random strategy targeting at the identified GUI widgets. The operations and the target GUI widgets are respectively chosen with a random algorithm. The GUI widgets are determined first because some widgets only support some actions. For example, an \texttt{input()} operation cannot be applied on a \texttt{Button}. Specifically, the \texttt{scroll()} does not require a specific target widget, and is directly conducted on the GUI screen. However, due to the natural feature of the robotic arm, extra time overhead is inevitable. We design a PoP-guided exploration strategy to improve exploration efficiency. Generally, the robotic arm starts from one specific point of the GUI screen. According to the least surprise principle \cite{geoffrey1987law, peter2001cranky, eric2003applying} in the field of GUI design, GUI widgets have spatial locality, which means that widgets with functionality relations are spatially adjacent in the GUI layout \cite{li2022cross}. Therefore, \toolname follows the Principle of Proximity (PoP) to determine the next target widget based on previous operation locations. The last two movements of the robotic arm are recorded as historical information that determines the exploration direction. When the robotic arm reaches the device edge, another randomly selected direction will guide the exploration.

Beyond the crash bugs, some compatibility bugs are hard to detect by intrusive frameworks. Such bugs are introduced by the irregular GUI screens. Some apps may not fit the special screen shape, and some widgets can be overlapped by the irregular part of the GUI screen, like a camera in a rounded rectangle area. For intrusive approaches, the overlapped widgets can be directly obtained by the ID or the XPath, but from the perspective of the app end user, such widgets are inaccessible. Under such a circumstance, we design a GUI comparison mechanism to compare the GUI responses to test operations between devices with regular and irregular GUI screens. Consequently, \toolname, as a non-intrusive approach, can effectively detect compatibility bugs by GUI comparison.

To evaluate the effectiveness and efficiency of \toolname, we design and implement an empirical evaluation on 20 real-world mobile apps. The results show that \toolname can effectively and efficiently explore the apps and detect bugs. We also conduct a case study on a representative industrial embedded device, showing that \toolname is with well-designed generalizability.

To sum up, the noteworthy contributions of this paper can be concluded as follows:

\begin{itemize}
	\item We propose a set of GUI screen and GUI widget identification technology designed for the robotic arm scenario to improve the generalizability of \toolname.
	\item We design a set of robotic arm movements that can flexibly and effectively simulate human testing operations, further attached with a PoP-guided exploration strategy to improve the GUI exploration efficiency.
	\item We introduce a novel comparison-based GUI compatibility bug detection method that is practical in detecting bugs beyond crashes.
	\item We implement a practical tool and conduct an experiment on real-world apps and a case study on a representative industrial embedded system to illustrate the effectiveness, efficiency, and generalizability of \toolname.
\end{itemize}

\textbf{The reproduction package and more details can be found from \underline{\url{https://sites.google.com/view/robotest2023}}.}

\section{Motivation}

\subsection{Dilemma of Intrusive Frameworks}

In this section, we provide two real-world examples to illustrate the motivation. The first motivation is a compatibility bug triggered by the irregular GUI screens of mobile devices. The second example is an embedded system example that does not support current intrusive GUI testing frameworks.

\subsubsection{Irregular Screen Motivating Example}

\begin{figure}[!htbp]
\centering
\includegraphics[width=\linewidth]{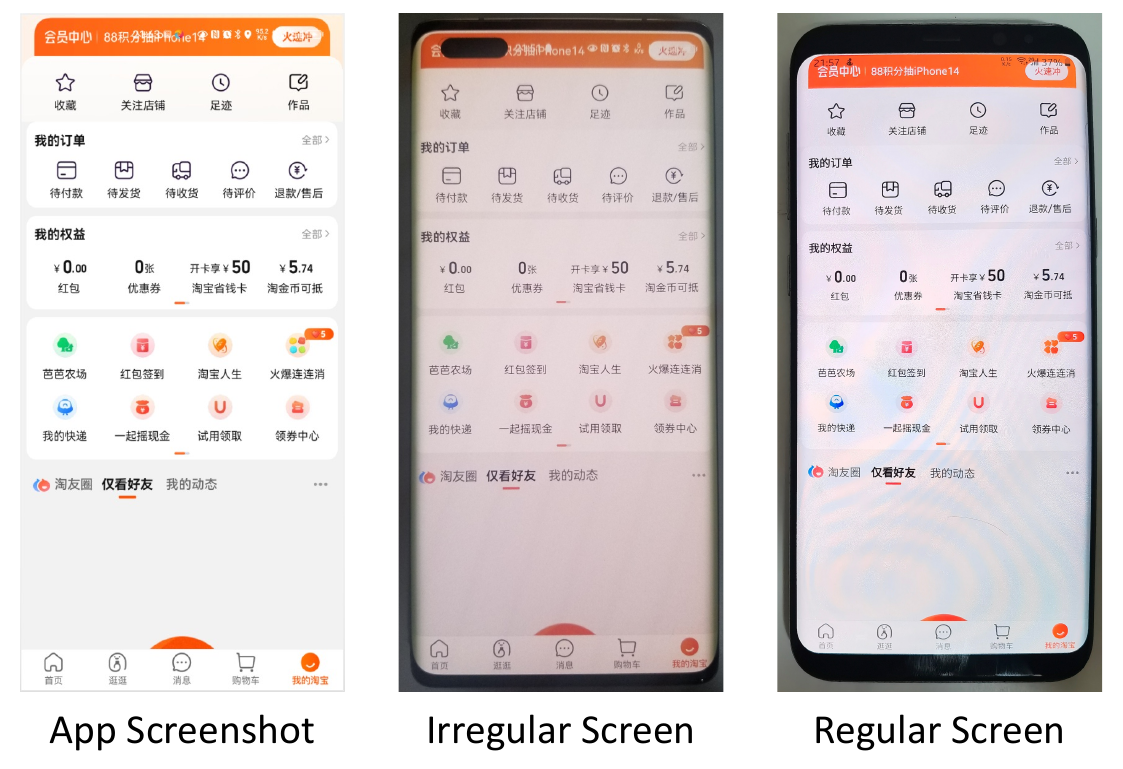}
\caption{A Compatibility Bug Triggered by Irregular Screen}
\label{fig:screen}
\end{figure}

In the first motivating example, we provide a typical compatibility bug that is triggered by the irregular GUI screen. 
The left image of \figref{fig:screen} is a screenshot of an activity for a famous shopping app, \textit{Taobao}, with over 500 million users. The app screenshot (\figref{fig:screen} left) does not show any bugs. However, with the comparison of the middle and right images of \figref{fig:screen}, which are respectively the corresponding photos taken for the specific app activity on a device with an irregular GUI screen and on a device with a regular GUI screen. It is easy to find that the camera of the mobile phone overlaps one button in the top-left corner of the app activity.

For traditional intrusive approaches, they may directly retrieve such a button from the GUI layout file of the app activity. For some more advanced approaches, they may directly recognize the button widget with CV technologies being adopted on the app screenshots, which are directly retrieved with specific methods provided by the intrusive GUI testing frameworks. However, no matter what strategies the traditional intrusive approaches take, the overlapped button widget will not be recognized because such a bug can only be detected from the actual user perspective. 

Existing intrusive approaches cannot effectively perceive the AUTs exactly as end users and biases exist when intrusive GUI testing frameworks simulate the AUT activity capture via the interface invoking. Therefore, a non-intrusive approach is necessary that views AUTs from the perspective exactly as end users.

Regarding the operation aspect, the intrusive GUI testing frameworks can easily operate on the widgets that are overlapped by the irregular GUI screens, because they use widget identifiers\footnote{\url{https://appium.io/docs/en/2.0/intro/clients/}} (\eg ID, XPath) to identify the target GUI widgets and the operations are directly invoked and applied to the target widgets of specific identifiers. However, such operation simulation of intrusive GUI testing frameworks brings bias as well. From the perspective of real-world end users, such target widgets are not accessible so they cannot be operated. Such a compatibility bug will be neglected by the intrusive GUI testing frameworks.

\subsubsection{UOS Motivating Example}

Ubiquitous operating systems (UOS) \cite{weiser1991computer} have risen as an extension of cyber-physical systems by adding human beings. A typical example is intelligent manufacturing systems. XiUOS is a representative UOS instance \cite{mei2022ubiquitous, cao2022xiuos}. XiUOS is designed to run on resource-limited Internet of Things (IoT) devices deployed in an industrial environment. 

\begin{figure}[!htbp]
\centering
\includegraphics[width=\linewidth]{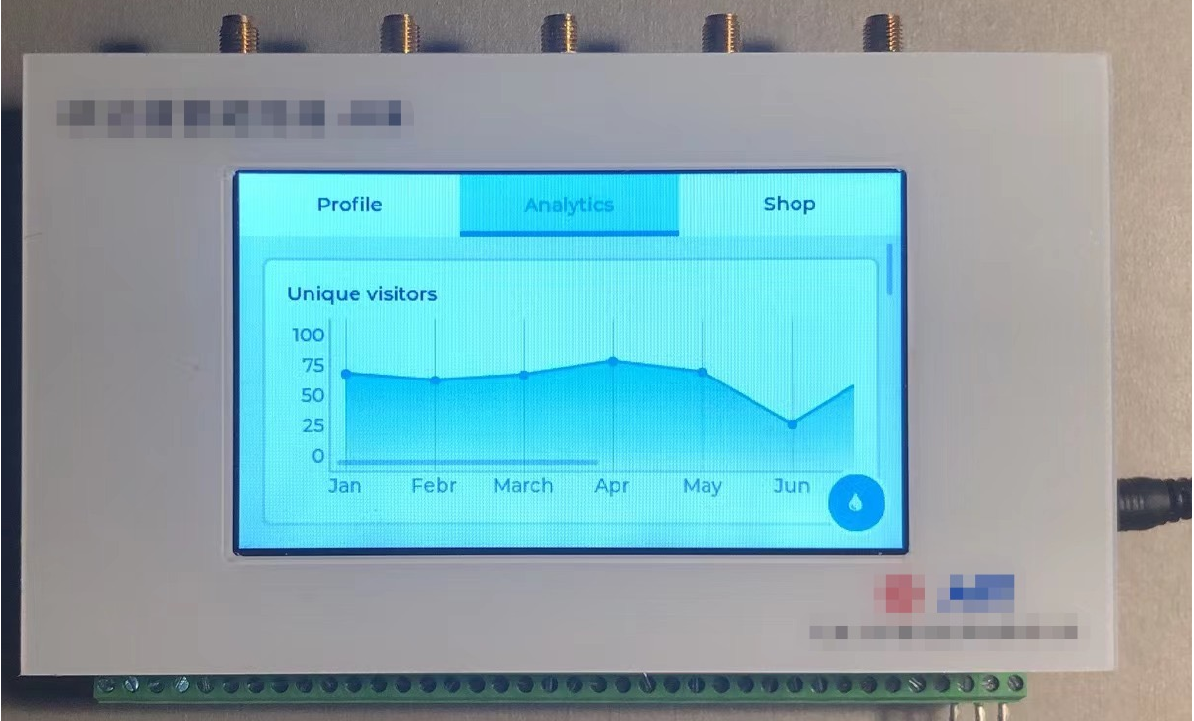}
\caption{A Representative UOS System: XiUOS}
\label{fig:xiuos}
\end{figure}

For the UOS represented by XiUOS (\figref{fig:xiuos}), the application quality insurance remains a challenge, because the applications are highly integrated on hardware devices, and code-level white-box testing is not enough to fully simulate the performance in runtime environments and real-world interactions \cite{cao2022xiuos}. It is important to apply GUI testing to the XiUOS application quality insurance. GUI testing starts from the end-user perspective and helps validate the compatibility of diverse GUI widgets and GUI operations. However, the challenge is that there is no testing framework supporting GUI testing on XiUOS devices because the hardware cannot provide the required interfaces. Moreover, the domain-specific nature of ubiquitous computing scenarios leads to a large number of customized XiUOS systems that are developed and widely applied instead of a unified universal one. Considering the non-intrusive nature and the diverse scenario characteristics, non-intrusive GUI testing with the visual-based robotic arm should be a practical solution.

\subsection{Limitations of Existing Non-intrusive Frameworks}

Existing non-intrusive GUI testing frameworks \cite{ran2022automated, qian2020roscript} have preliminary explore the potential capability of robotic arms in mobile app GUI testing. Their efforts and the aforementioned challenges have shown the importance of non-intrusive GUI testing. However, existing non-intrusive GUI testing frameworks show some limitations in being applied to practical GUI testing tasks.

First, existing non-intrusive frameworks do not adopt adaptive GUI widget and screen detection. These frameworks fix the devices, requiring extra configurations on the device screen area for taking photos. This will hugely reduce the practicality of the non-intrusive frameworks. Moreover, existing non-intrusive frameworks do not involve the necessary processing of the photos taken by the robotic arm cameras and do not adjust existing CV algorithms to fit the physically taken photos instead of app screenshots. It is challenging to both detect GUI screens out of backgrounds from the physically taken photos and detect GUI widgets from the detected screens. In order to distinguish the device screens from the physical background to avoid misrecognition without manual labor, we adjust the CV algorithms and adopt an adaptive GUI widget and screen detection for different device screens. Moreover, we design a set of specially adjusted GUI widget and screen detection algorithms for the target detection in physically taken photos. The design of \toolname is supposed to eliminate the potential influences of physical conditions. 

Second, existing non-intrusive frameworks cannot faithfully simulate user operations. The XY-plane robotic arms used by existing non-intrusive frameworks are not flexible and restrict the movements of the robotic arm on a plane, which cannot realistically simulate the actual behaviors of human testers. Moreover, such XY-plane robotic arms will lead to more movement overhead, as shown in our experiment results in Table 1. In \toolname, we use a 4-DOF robotic arm to simulate user operations of human testers more flexibly, and it can directly move to the target widget location exactly as human testers do to reduce movement overhead, working together with our PoP-guided strategy. Therefore, we believe that the 4-DOF robotic arm is a better choice for mobile app GUI testing.

Third, existing non-intrusive frameworks do not involve the non-intrusive detection of compatibility bugs. Actually, there are many compatibility bugs that cannot be found from the pure software perspective \cite{ren2022cross}, only considering the different features within intrusively obtained screenshots of app GUI by inner interface invoking, which cannot realistically reflect the app running situation from the user perspective. Instead, \toolname considers the externally taken photos containing app GUI and device features, \eg irregular screen. Detecting compatibility bugs via the combination of app GUI and device features is more realistic and effective. Compatibility bugs occur when the apps do not fit the hardware devices, which is common in mobile app testing. We give examples in \figref{fig:screen} and \figref{fig:result}. This is why non-intrusive frameworks are necessary. In our work, \toolname supports compatibility bug detection by the GUI analysis and comparison. \toolname helps find some compatibility bugs in the experiment.

\section{Approach}

\begin{figure*}[!htbp]
\centering
\includegraphics[width=\linewidth]{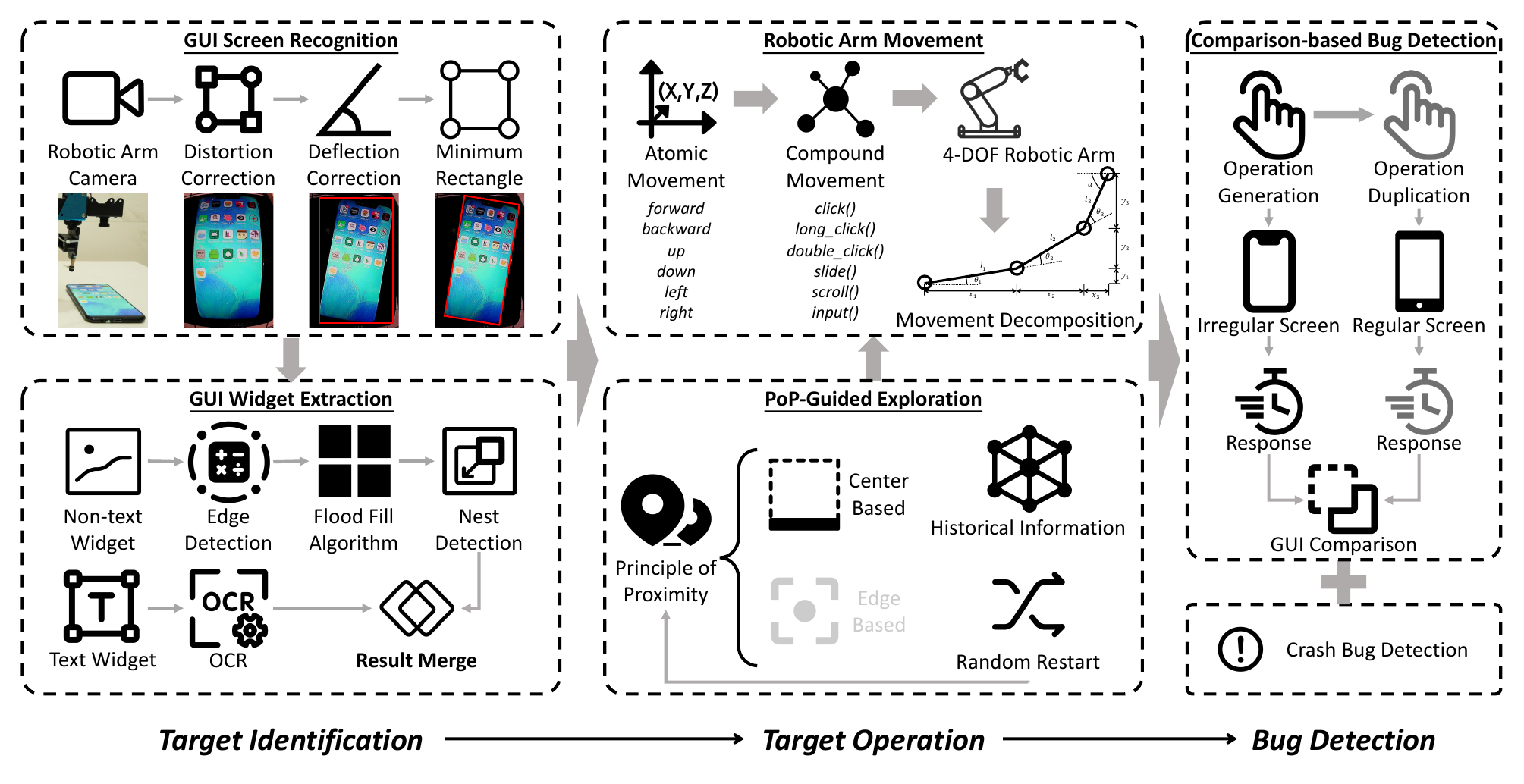}
\caption{\toolname Framework}
\label{fig:framework}
\end{figure*}

\toolname is a practical non-intrusive GUI exploration testing approach with visual-based robotic arms. The overall framework is shown in \figref{fig:framework}. For the target identification, \toolname involves the GUI screen recognition and the GUI widget extraction with the combination of traditional CV technologies and advanced DL models. In order to fit the requirements of different granularities (\ie GUI screens and GUI widgets), we fine-tune the algorithms with different parameters and train the models with different datasets. For the target operation, we design a complete set of compound robotic arm movement instructions based on the atomic movements (forward, backward, up, down, left, and right). Considering the overhead of robotic arm testing operation, we design a PoP-guided strategy to improve robotic exploration efficiency. The main purpose of \toolname is to non-intrusively detect bugs that are closely related to the hardware features, we also introduce a GUI comparison-based compatibility bug detection mechanism, which can duplicate the operations to a reference device with the robotic arm and compare the GUI similarity of the subsequent GUI images. According to the similarity, which reflects the robotic arm operation responses, \toolname can detect compatibility bugs.

\subsection{GUI Screen Recognition}

Different from the intrusive approaches, where the screen is directly accessible and does not need extra recognition with existing screenshot capture methods, the GUI screens need to be recognized from the physical photos in the non-intrusive GUI testing scenario. To recognize the GUI screen, we first adopt the CV algorithms to identify the approximate GUI screen, which is assumed as a rectangle-like area. The specific algorithm is the edge detection and the flood fill algorithm \cite{yu2021layout, chen2020object}. Then, we can obtain the approximate area of the GUI screens. However, the area is not exactly the GUI screen, so we have further processing.

To discover the GUI screen from a physical space, there are two main challenges. The first is image distortion, which is caused by the angles of the robotic arm camera, and any offset from directly above the device will lead to distortion. The second challenge is caused by the relative location between the device under test and the robotic arm. If there is any deflection on the placement of the device, the image will have an angle, which should be corrected.

\begin{equation}
\label{form:distortion}
\begin{bmatrix}
f_x & s & c_x \\ 0 & f_y & c_y \\ 0 & 0 & 1
\end{bmatrix}
\end{equation}

To address the first challenge regarding image distortion, we adopt the generalizable camera calibration algorithm \cite{zhang2000flexible}, which is a fully-automated process. Specifically, we use a chessboard image as a template, following the common practice. The advantage of the chessboard image is that it has very clear lines with black-and-white boxes. Then we take photos on the template chessboard image with the robotic arm camera from different angles. According to the geometric imaging model of the robotic arm camera, we can calculate the distortion according to \formref{form:distortion}, where $[c_x, c_y]$ refers to the optical center (the principal point) in pixels, $[f_x, f_y]$ refers to focal lengths in pixels, and $s$ refers to the skew coefficient, which is non-zero if the image axes are not perpendicular.

Regarding the second challenge of GUI screen recognition, we first detect the image paralleling rectangle that can contain the deflected GUI screen, and then we can obtain the coordinates of the intersection points of the GUI screen and the photo taken by the robotic arm camera. Then, the deflection angle can be calculated. Meanwhile, the scale between the paralleling rectangle area and the actual GUI screen size can also be calculated. The deflection angle and scale values are stored for the movement transformation.

After the distortion correction and the deflection correction on the initial recognized GUI screen area, we can obtain a precise GUI screen recognition result.

\subsection{GUI Widget Extraction}

GUI widget extraction from the visual perspective has been widely used in several GUI testing tasks \cite{yu2021layout, chen2020object, yu2021prioritize}. However, the app screenshots under analysis are with different qualities in existing studies and in this work. The screenshots taken directly with methods of intrusive frameworks are with high resolution, but images taken by the robotic arm camera are with low resolution and may be affected by the light conditions of physical space. Therefore, we have special processes beyond existing studies, to achieve better widget extraction effectiveness in the non-intrusive scenario. All widgets in the app GUI are divided into two categories \cite{chen2020object} according to the features, the text widgets and non-text widgets. Then we adopt different algorithms for different widget types. 

For the text widgets, we use the OCR algorithms. Considering the feature of images taken by the robotic arm camera, we adopt a two-granularity OCR extraction strategy, which aims at improving extraction effectiveness. Specifically, we first roughly extract text widgets with OCR to the app GUI image as a whole. The obtained information includes all the text widgets, attached with their concrete texts, widget coordinates, and text extraction confidence values. However, due to the low-resolution nature and potential disturbance of physical space, we conduct a second refined extraction to the results of the first rough extraction. The refined extraction focuses on the first rough extraction results and applies the OCR algorithm once again. In the refined extraction, we can obtain each word or character in the sentences extracted in the first step, and the results with extraction confidence lower than a threshold will be eliminated. The threshold is set as 0.8 based on existing studies \cite{yu2021layout, chen2020object} and a small-scale pilot study. Besides, we calculate the space between the extracted words and characters and make comparisons. The abnormal results will be merged if the calculated space is smaller than others. Then, we can obtain the text widgets with better extraction effectiveness in the non-intrusive scenario.

For the non-text widgets, we first apply the edge detection algorithm, specifically Canny \cite{canny1986computational}, to extract all existing edges in the images taken by the robotic arm camera. Then, we conduct the morphological closing operations to connect as many broken edges as possible. For each connected contour of the extracted edges, we use a minimum coverable rectangle to label it. Such contours are considered as widget candidates. Further, we adopt two rules \cite{yu2021layout, chen2020object} to eliminate mis-extracted candidates. First, for the nested candidates, we iteratively use one or more nested candidates to replace the nesting candidates if the sizes of the nested candidates are larger than one preset threshold. The threshold is set as 0.85, following the practice of existing studies \cite{yu2021layout, chen2020object}. Second, for the candidates with an extremely large size to the whole app GUI, we will eliminate them, following the common practice \cite{yu2021layout}.

Finally, we merge the extracted text and non-text widgets according to their coordinates to remove duplicates.

\subsection{Robotic Arm Movement}

Robotic arm movements are designed to simulate the testing operations of human testers. Existing non-intrusive approaches adopt the XY-plane robotic arms \cite{ran2022automated, qian2020roscript}. However, in such practices, the movements of the robotic arm are restricted on a plane, which cannot realistically simulate the actual behaviors of human testers. Therefore, we use a robotic arm with four degrees of freedom, which can more realistically simulate operations in a three-dimension physical space. 

\begin{figure}[!htbp]
\centering
\includegraphics[width=\linewidth]{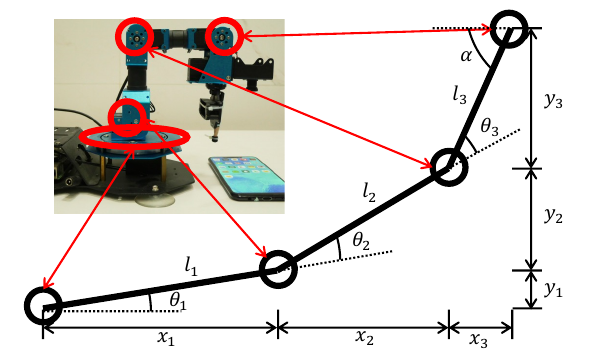}
\caption{Movement Decomposition to the 4-DOF Robotic Arm}
\label{fig:arm}
\end{figure}

During the GUI testing process, all the test operations can be viewed as the compound of atomic operations, which is the movement of one single direction with a specific distance. In order to uniformly represent the movement of the robotic arm of 4-DOF, we design in total 6 atomic movements, \ie forward, backward, up, down, left, and right.

Since the robotic arm has 4-DOF, we need to decompose the movements to different degrees of freedom. The movements of different DOFs are implemented as the rotation of different DOFs. Therefore, the changes of $\theta_1$, $\theta_2$, and $\theta_3$ determine the changes of $x_1$, $x_2$, $x_3$, $y_1$, $y_2$, and $y_3$. 
The robotic arm is placed in a 3-dimension XYZ-coordinate system. In \figref{fig:arm}, we show the simplified robotic map on the XY-plane as an example, and the decomposition is the same on the XZ-plane and the YZ-plane. The $\theta_i$ refers to the angles between different sticks. The $x_i$ and $y_i$ refer to the projection length of different sticks with length $l_i$. In order to determine the movement of the robotic arm from point $A(x_a, y_a)$ to point $B(x_b, y_b)$ (as a projection on the XY-plane), we need to calculate the change $(\Delta x, \Delta y)$, which is determined by the $(\Delta \theta_1, \Delta \theta_2, \Delta \theta_3)$, so we need to calculate the static angles $\theta_i$ at specific points. For the robotic arm at point $A$, where $x_a = x_1 + x_2 + x_3$ and $y_a = y_1 + y_2 + y_3$. The angle $\alpha = \theta_1 + \theta_2 + \theta_3$ shows the end of the robotic arm to the device under test. $l_1$, $l_2$, and $l_3$ respectively refer to the arm lengths. Then, the $A(x_a, y_a)$ can be calculated as:

\begin{subequations}
	\begin{align}
		x_a = l_1 cos(\theta_1) + l_2 cos(\theta_1 + \theta_2) + l_3 cos(\theta_1 + \theta_2 + \theta_3)
	\end{align}
	\begin{align}
		y_a = l_1 sin(\theta_1) + l_2 sin(\theta_1 + \theta_2) + l_3 sin(\theta_1 + \theta_2 + \theta_3)
	\end{align}
\end{subequations}

Then, we can obtain the values of $\theta_1$, $\theta_2$, and $\theta_3$ as:

\begin{subequations}
	\begin{align}
	\begin{split}
		\theta_1 = &arctan(\frac{y_1 + y_2}{x_1 + x_2}) -\\ &arccos(\frac{(x_1 + x_2)^2 + (y_1 + y_2)^2 + l_1^2 - l_2^2}{2 l_1 \sqrt{(x_1 + x_2)^2 + (y_1 + y_2)^2}})		
	\end{split}
	\end{align}
	\begin{align}
		\theta_2 = arccos(\frac{(x_1 + x_2)^2 + (y_1 + y_2)^2 - l_1^2 - l_2^2}{2 l_1 l_2})
	\end{align}
	\begin{align}
		\theta_3 = \alpha - \theta_1 - \theta_2
	\end{align}
\end{subequations}

With the atomic movements, we design compound movements, which actually simulate the testing operations of human testers. \toolname involves six most widely-used actions of human testers, including \texttt{click()}, \texttt{double\_click()}, \texttt{long\_click()}, \texttt{slide()}, \\ \texttt{scroll()}, and \texttt{input()}. Three operations: \texttt{click()}, \texttt{double\_click()}, and \texttt{long\_click()}, adopt a similar strategy, which is to find the location of the target widget and move the robotic arm to right above the target. Then, the \texttt{click()} operation attaches the down-and-up operation pair once, the \texttt{double\_click()} operation attaches the down-and-up operation pair twice in a short time, the \texttt{long\_click()} attaches the down-and-up operation pair and have a short stay. For the \texttt{slide()} operation, \toolname focuses on a specific widget and drags the widget from one edge to another. For the \texttt{scroll()} operation, \toolname does not focus on specific widgets, but directly scrolls the page from the bottom to the top, or vice versa. For the \texttt{input()} operation, \toolname can simulate what human testers do, which means it first clicks on the target input area and then inputs the characters of the input text one by one through the soft keyboard of the device under test. The identification of a soft keyboard is like the text widget detection part of \toolname. During the operation, the deflection angle and scale values are referred to for the movement transformation.

\subsection{PoP-Guided Exploration Strategy}

Due to the extra time overhead brought by the robotic arm movement, efficiency is lowered for the non-intrusive GUI exploration testing approach. In order to improve the exploration efficiency and keep the exploration effectiveness, we propose a PoP-guided exploration strategy. During the design of the PoP-guided exploration strategy, our purpose is to explore as many app GUI states as possible with the shortest robotic arm movement (\ie time overhead). Inspired by the principle of least surprise in GUI design, \ie spatial locality and responsive patterns \cite{seebach2001cranky, raymond2003applying, li2022cross}, and referring to the Principle of Proximity (PoP), we adopt the center-based PoP-guided exploration strategy. Following the Principle of Proximity, \toolname determines the next target widget based on the previous operation locations. The last two operations of the robotic arm are recorded as historical information that determines the exploration direction. When the robotic arm reaches the device edge, another randomly selected direction will guide the exploration. This practice is considered to effectively shorten the movement distance of the robotic arm and improve exploration efficiency.

\subsection{Comparison-based Bug Detection} 

\begin{figure}[!htbp]
\centering
\includegraphics[width=\linewidth]{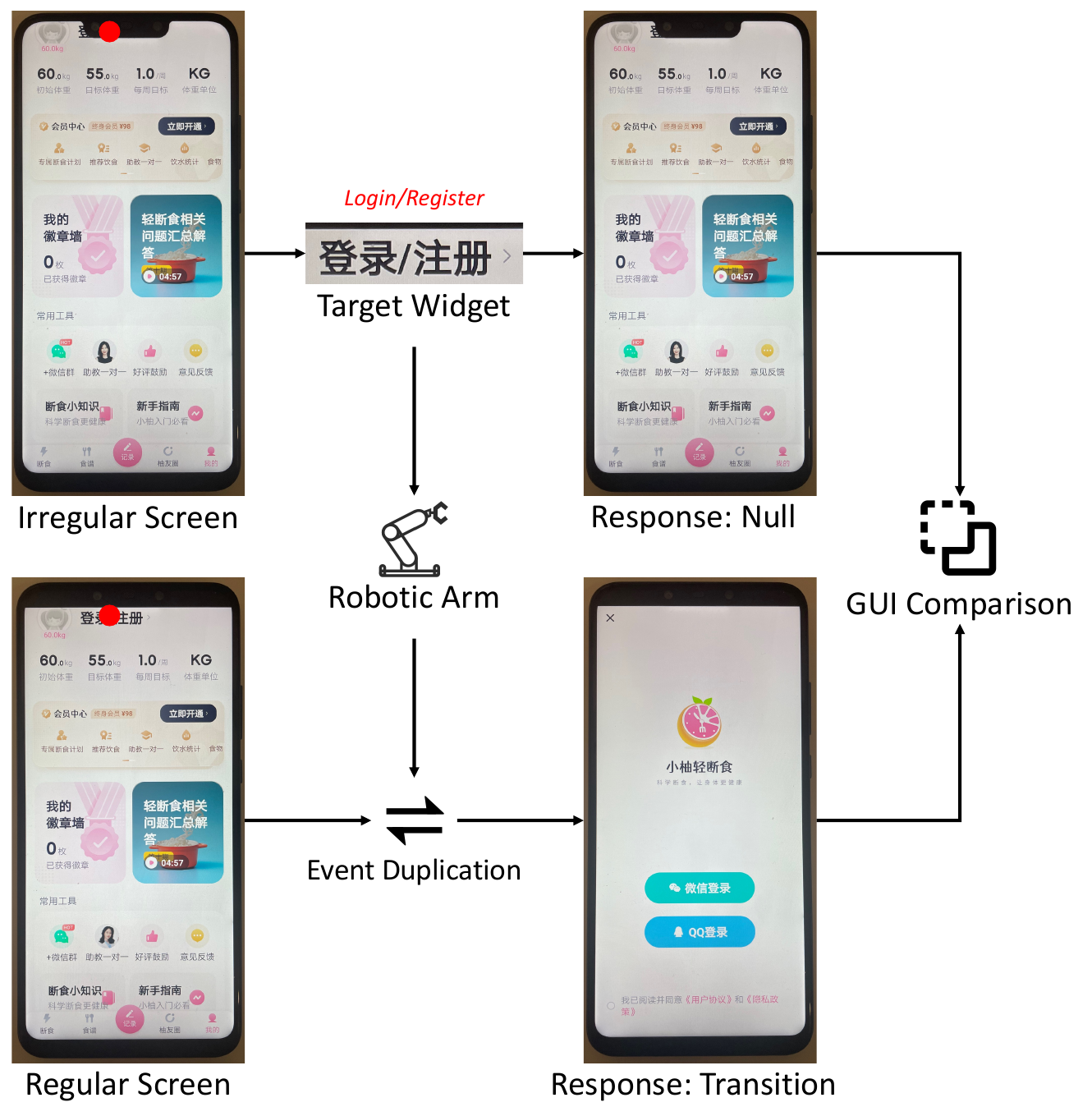}
\caption{Comparison-based Compatibility Bug Detection}
\label{fig:detection}
\end{figure}

Besides the crash bugs, we also focus on the compatibility bugs triggered by the device features, like the GUI screen shapes. For example, on devices of irregular GUI screen shape (\figref{fig:detection}), the widgets may be overlapped by the camera and sensor area, leading to the inaccessibility of such widgets. However, oracles of such compatibilities are hard to be automatically generated if we only focus on the responses of specific operations. The reason is that other factors may lead to similar responses. For example, if the robotic arm clicks on a text widget, which is correct to cause no responses, just like clicking on the widgets in the overlapped area. Some existing studies \cite{choudhary2010webdiff} \cite{fazzini2017automated} propose approaches to identify compatibility bugs. However, such approaches are intrusive and cannot be applied to the robotic arm-based testing scenario.

In order to automatically detect such compatibility bugs, we have an investigation into such bugs, and we find that such bugs are caused by the incompatibility between hardware features and GUI designs, which means that such compatibility bugs will not happen on devices without special hardware features. Therefore, we design a GUI comparison-based compatibility bug detection approach. Specifically, we use two devices to compare the responses and GUI changes. The device with hardware features (\ie irregular GUI screen shape) is used to generate test operations, then the operations are duplicated to a reference device with the robotic arm, which is without hardware features (\ie regular GUI screen shape). We ensure that the two devices are with the same screen size so that the GUI widgets can stay in the same layout. Then we observe the responses and GUI changes. If the responses on both devices happen, we can infer that the operations are normal. If the responses on both devices do not happen, we can infer that it is likely that the operation is conducted on a non-clickable text widget, which is also a normal operation, though without any testing progress. If the responses on two devices are different, there should be a compatibility bug. The GUI comparison is calculated by the SIFT algorithm \cite{lowe1999object} on GUI images. Like the example we give in \figref{fig:detection}, on the first device, the operation is generated on a widget overlapped by the camera and sensor area, so the response is null. However, the operation duplicated to the reference device causes an activity transition. The difference between the app GUI indicates the occurrence of a compatibility bug.

\section{Empirical Evaluation}

\begin{table*}[!htbp]
\centering
\caption{Empirical Evaluation Result Summary of \toolname}
\scalebox{0.9}{\begin{tabular}{ccc||r|r||r|r||r}
\toprule
Approach & 
Type & 
Granularity 
& Line Coverage (\%)
& Branch Coverage (\%)
& Crash Bug
& Compatibility Bug
& Distance (mm)
\\ \midrule

\multirow{2}{*}{\monkey} & \multirow{2}{*}{Intrusive} 
  & 3600s   & 32.23\% & 20.15\% & 151 & - & N/A \\ 
& & 200step & 24.47\% & 13.28\% & 118 & - & N/A \\ \midrule

\multirow{2}{*}{\vtest}  & \multirow{2}{*}{Non-Intrusive} 
  & 3600s   & 26.55\% & 15.20\% & 105 & - & 1486.30 \\ 
& & 200step & 26.71\% & 15.32\% & 78  & - & 1470.37 \\ \midrule

\multirow{2}{*}{\toolrandom} & \multirow{2}{*}{Non-Intrusive} 
  & 3600s   & 29.04\% & 18.48\% & 120 & 6 & 1439.00 \\  
& & 200step & 29.50\% & 18.77\% & 102 & 7 & 1517.55 \\ \midrule

\multirow{2}{*}{\tooledge} & \multirow{2}{*}{Non-Intrusive} 
  & 3600s   & 29.62\% & 19.41\% & 126 & 9 & 1173.65 \\  
& & 200step & 30.44\% & 19.86\% & 107 & 8 & 1201.95 \\ \midrule

\multirow{2}{*}{\toolcenter} & \multirow{2}{*}{Non-Intrusive} 
  & 3600s   & 30.23\% & 20.16\% & 128 & 13 & 1062.70 \\
& & 200step & 31.13\% & 20.38\% & 108 & 12 & 1096.55 \\ \bottomrule

\end{tabular}}
\label{tab:exp}
\end{table*}

\subsection{Evaluation Setting}

\subsubsection{Research Question (RQ)}

We focus on three research questions, which respectively target at the code coverage, bug detection capability, and time overhead of \toolname. The research questions pay attention to both effectiveness and efficiency aspects.

\textbf{RQ1 (Code Coverage)}: How effective is \toolname in code coverage compared with baselines and different exploration strategies?

\textbf{RQ2 (Bug Detection)}: How effective is \toolname in detecting real-world bugs compared with baselines and different strategies?

\textbf{RQ3 (Time Overhead)}: What is the time overhead of \toolname compared with baselines and different exploration strategies?

\subsubsection{Experiment Subject}

We use 20 real-world mobile apps as the experiment subjects. Such apps are used in existing studies like \cite{ran2022automated, su2017guided, pan2020reinforcement}. We use the apps that are still usable for our evaluation. The running environment is on a Huawei Mate 20 device with Android 9 for all baselines and the proposed approach. In order to capture the coverage and bug information, we instrument such apps with Jacoco\footnote{https://www.jacoco.org/jacoco/}, which is widely-used in academic research on app exploration testing. Regarding the code coverage, the target is the whole app, and we use the widely used tool Jacoco to calculate the coverage information, including line coverage and branch coverage. The criteria of identifying and confirming distinct bugs are based on the activity information and stack trace logs. One thing to notice is that the instrumentation is only for evaluation purposes, and it will not break the non-intrusive feature of the proposed \toolname. We put the detailed information of the experiment subject apps on our online package due to the page limit.

\subsubsection{Baseline Approach}

To evaluate the effectiveness of \toolname, we use three groups of baseline approaches. The first baseline is \monkey \cite{monkey}, which is the most widely-used Android exploration testing tool, and has been integrated into the Android development environment. \monkey is an intrusive approach. The second is \vtest \cite{ran2022automated}, proposed by Ran \etal. It is the state-of-the-art and representative approach for non-intrusive GUI exploration testing. Another non-intrusive framework, RoScript \cite{qian2020roscript}, is not compared because: 1) RoScript adopts the same widget detection algorithms as \vtest; 2) RoScript adopts the same robotic arm and movement strategy as \vtest; 3) \vtest can be considered as a combination of a random exploration algorithm and the RoScript, which is originally for test script record and replay. Similarly, other related studies listed in this paper on visual-based GUI testing approaches are designed for different software testing tasks, like report reproduction, report deduplication, \etc They cannot be used for the baselines of our work. Another group of baselines is the variants of \toolname, which we label as \toolrandom and \tooledge. For the actual \toolname, we label it as \toolcenter. The variants are related to the PoP-guided exploration strategy. For \toolrandom, we remove the exploration strategy and conduct random exploration based on the recognized widgets. For \tooledge, we adopt a similar exploration strategy like \toolcenter but move the starting point to the edge of the device under test. The experiments for all the baselines and variants of \toolname are repeated 10 times to eliminate the potential effect of randomness. The source code of the baseline approaches are not fully provided. Therefore, we ensure that the baseline approaches are faithfully implemented by referring the original paper description, open-source repositories, and the corresponding algorithms they adopt. We also completely simulate the XY-plane robotic arm they use.

\subsubsection{Evaluation Metric}

We totally use three kinds of evaluation metrics, corresponding to the three research questions. For code coverage, we use the widely-used line coverage and branch coverage. For bug detection capability, we focus on crash bugs and compatibility bugs, respectively. For the time overhead assessment, we use the movement distance instead of directly calculating the time because the distance is purely related to the robotic arm movement, which is what we really focus on considering the extra overhead brought by using the robotic arm to conduct non-intrusive GUI testing. In the experiment, we conduct the experiments on the same device to eliminate the potential influence of device screen sizes on the result comparison. The movement distance and time overhead are positively correlated linearly because the speed of the robotic arm is constant. To calculate the metric values, we run each approach for 3600 seconds. Besides, in order to better show the effectiveness, we compare with the same number of test steps (200).

\subsection{RQ1: Code coverage}

\begin{figure*}[!htbp]
\centering
\includegraphics[width=\linewidth]{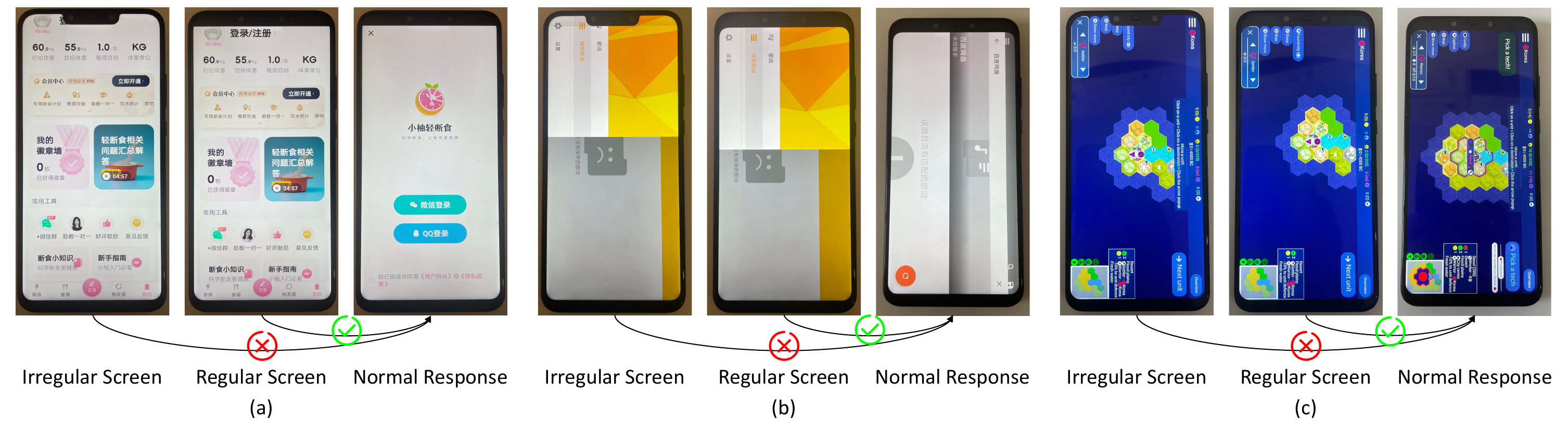}
\caption{Examples of Detected Compatibility Bug}
\label{fig:result}
\end{figure*}

The first research question focuses on a traditional effectiveness evaluation metric, coverage. In order to more comprehensively evaluate the coverage, we use the two most widely-used coverage: line coverage and branch coverage. The results are shown in \tabref{tab:exp}. Compared with the existing state-of-the-art non-intrusive baseline approach, \toolcenter outperforms \vtest by 13.9\% and 32.7\% on line coverage and branch coverage, respectively, for the 3600-second running. For the 200-step result, \toolcenter outperforms \vtest by 16.6\% and 33.0\% on line coverage and branch coverage. Moreover, \toolrandom, which is the no-guidance-strategy variant of \toolname, outperforms \vtest by 9.4\% -- 22.5\% on different coverage and test operation granularities. The results show that as a non-intrusive approach, \toolname can perform much better than the state-of-the-art approach. The advantage is not only due to the specially designed GUI screen recognition and GUI widget extraction but also due to the introduction of the PoP-guided exploration strategy, which can improve the testing efficiency. Therefore, within a fixed time period, \toolname can generate more test operations to achieve more code coverage. We also compare \toolname with \monkey, which can generate test events with extremely high efficiency. Though \toolname performs similarly in line coverage, it performs better in branch coverage. The reason is that the target of \toolname is the extracted widgets, but \monkey generates test events based on the GUI image pixels. This further illustrates the effectiveness of the GUI screen recognition and GUI widget extraction design of \toolname.

\subsection{RQ2: Bug Detection}

In order to evaluate the bug detection capability of \toolname and baselines, we especially focus on compatibility bugs. In total, we find 13 and 12 different compatibility bugs in the 3600-second results and 300-step results, respectively. Such bugs cannot be found by the \monkey and \vtest baselines. \monkey is an intrusive approach, the compatibility bugs are completely ignored. To validate the detected bugs, we invite testing experts to manually re-confirm all the detected bugs, including crashes and compatibility bugs. Based on the manual check, we confirm that all the bugs reported in this paper are valid and unique.

To better illustrate the found real-world compatibility bugs in our evaluation, we give three examples in \figref{fig:result}. 
Example (a) is found in a diet assistant app. The ``Login/Register'' button is overlapped by the camera and sensor area. When the robotic arm clicks on this button on different devices, the app will respond differently, and on a regular GUI screen, the app transits to the Login Activity.
Example (b) is found in a music app. The ``Lyric'' button is overlapped by the camera and sensor area of the irregular area. When the button is clicked by the robotic arm, the app will not transit to the Lyric Activity as expected on the irregular GUI screen. 
Example (c) is found in a video game app. The ``Sleep'' and ``Found City'' buttons are overlapped by the camera and sensor area of the irregular area. The behavior is similar to the above two examples, and the operation on such widgets will lead to different responses on different devices.

Besides, we calculate the total number of detected crash bugs. Compared with \vtest, \toolcenter can find 23 and 30 more crash bugs in 3600-second results and 300-step results, respectively. \toolcenter also finds 8 and 6 more bugs than \toolrandom and 2 and 1 more bugs than \tooledge. The results show that the PoP-guided exploration strategy can improve testing efficiency and find more bugs. 

\subsection{RQ3: Time Overhead}

As a primary limitation of non-intrusive approaches, efficiency is another focus. Considering the major difference between traditional intrusive and non-intrusive approaches, we use the robotic arm movement distance as the evaluation metric. When compared with the state-of-the-art non-intrusive approach, as shown in \tabref{tab:exp}, the \tooledge and \toolcenter have 21.0\% and 28.5\% shorter distances in the 3600-second results than \vtest. For the 200-step result, \tooledge and \toolcenter have 18.3\% and 25.4\% shorter distances in the 3600-second results than \vtest. 

In order to highlight the advantage of the proposed PoP-guided exploration strategy, \toolname is compared with different variants, the \tooledge and \toolrandom. Compared with \toolrandom, which completely removes the exploration strategy, \toolcenter has 26.2\% and 27.7\% shorter distances in 3600-second and 200-step results. Compared with \tooledge, which adopts a different exploration strategy, \toolcenter has 9.5\% and 8.8\% shorter distances in 3600-second and 200-step results. The results indicate that the PoP-guided exploration strategy can effectively improve the GUI exploration efficiency, and the center-based strategy is an optimized choice for non-intrusive GUI exploration testing.

\subsection{Threats to Validity}

The first threat to validity may be the experiment subject selection. We take some measures to alleviate such a threat. First, the apps we use are from representative existing studies, including \cite{ran2022automated, su2017guided, pan2020reinforcement}. Second, the used apps are very popular open-source apps, which have obtained a large number of \textit{stars} on their GitHub repositories. Third, the used apps cover different categories, which expands the generalizability of the proposed approach. 

The second potential threat may be the settings of used parameters or hyperparameters. \toolname involves some parameters as thresholds, \eg the similarity between GUI images for compatibility bug detection. However, we have followed the common practice from existing work to reduce the threat. Moreover, we find suitable settings with small-scale pilot studies. For the baseline approach implementation, we use the open-source package to avoid biases.

The third possible threat is how \toolname can actually simulate the test operations of human testers. Regarding the target (widget) identification, we use visual-based approaches to simulate human perspectives. Assume that with a strong visual algorithm, \toolname can effectively simulate the target identification of human testers. Regarding the target operation, Robotic arm movements of \toolname can fully simulate human operations with the 4-DOF design, which is flexible enough for human operation simulation. Briefly speaking, \toolname can theoretically simulate all human perspectives and find compatibility bugs.

The fourth threat might be the construction of baseline approaches. To minimize the potential threat to construct validity, we ensure that the baseline approaches are faithfully implemented by referring to the original paper description, open-source repositories, and the corresponding algorithms they adopt.

\section{XiUOS Case Study}

In order to illustrate the extensibility of \toolname in GUI exploration testing on the embedded systems, we have a case study on the representative XiUOS device, as shown in \figref{fig:xiuos}. We calculate the test operation generation effectiveness on two apps provided by LVGL\footnote{https://lvgl.io/}, the most popular open-source embedded graphics library. Apps on XiUOS are relatively less complex than mobile apps. Moreover, XiUOS does not support the instrumentation for code coverage calculation. Therefore, we manually count the covered pages and widgets as effective illustrations. Due to the uncontrollable feature of XiUOS apps, we run the GUI exploration testing with 50 steps. The results are shown in \tabref{tab:xiuos}.

\begin{table}[!htbp]
\centering
\caption{XiUOS Case Study Result}
\scalebox{1}{\begin{tabular}{crr}

\toprule           & App 1 & App 2 \\ \midrule
Effective Click    & 9     & 20    \\ \midrule
Noneffective Click & 27    & 15    \\ \midrule
Missing Target     & 1     & 2     \\ \midrule
Effective Slide    & 10    & 11    \\ \midrule
Noneffective Slide & 3     & 2     \\ \bottomrule

\end{tabular}}
\label{tab:xiuos}
\end{table}

For the first app, \toolname can correctly identify GUI widgets in 49 test operations and only one widget is misrecognized. The GUI widget recognition success rate reaches 98\%. For all the recognized GUI widgets, \toolname successfully generates test operations on 19 steps on different widgets, reaching 38\%. The failure reason is that this app contains some industrial machine operation data, which is not operable, and we count this situation as ``noneffective click and slide'' in \tabref{tab:xiuos}. On this app, we find a crash bug, which can be triggered by the click operation on one specific widget. We manually reproduce this bug and confirm it. The crash bug found on the first app is due to the memory leak, which can lead to the crash of the appv and even the no-response of the whole XiUOS device.
For the second app, \toolname effectively detects 96\% GUI widgets and misses two widgets. Regarding the operation generation successful rate, \toolname can effectively generate test operations in 62\% of all steps. The noneffective click and slide are mainly due to the images and texts. We also find a bug that a \textit{Switch} widget is not operable due to the compatibility of the app to the XiUOS device. The compatibility bug found on the second app is due to the compatibility issue between the device and the app, and the Switch widget is partly inaccessible on the device screen, making it not operable during the testing exploration.

We further calculate the time overhead on the GUI exploration testing case study on the XiUOS device. For a total of 100 test operations, the average photo taking is 1.16 seconds, the average GUI understanding and GUI widget extraction time is 2.05 seconds, the PoP-based guidance time is 0.69 seconds, and the robotic arm movement time is 2.47 seconds. The XiUOS system takes an average of 3.54 seconds to respond to testing operations. However, the response time varies a lot, the range is 0.1 seconds to over 8 seconds.

In summary, \toolname can effectively identify GUI widgets, generate testing operations, and find bugs on the XiUOS device, as a representative of embedded industry operating systems that cannot be intrusively tested. We believe the generalizability of \toolname is well elaborated, and it can be extended to more different scenarios, like the controlling screens on vehicles.

\section{Related Work}

\subsection{Robotic Arm in Software Engineering}

In recent years, robotic arms have been increasingly utilized in software engineering, especially in software testing, with the aim of meeting higher testing requirements and more complex testing scenarios.
Mao \etal propose Axiz \cite{mao2017robotic}, a robotic-based framework for automating mobile app testing with the aim of achieving truly black-box automation. 
Banerjee \etal \cite{banerjee2018robotic, banerjee2018image, banerjee2018object, banerjee2018integrated, banerjee2018hand} propose a series of integrated test automation frameworks for testing motion-based image capture systems using a robotic arm.
Banerjee \etal \cite{banerjee2018object} present a touch-to-track test automation framework. 
Banerjee \etal \cite{banerjee2018integrated} further propose an integrated test automation framework for testing motion-based image capture system.
Banerjee \etal \cite{banerjee2018hand} induct and program a robotic arm to simulate real-world user motions to enable HJR to simulate similar motions.
Maciel \etal \cite{maciel2022systematic} propose that allowing more realistic interactions is among the main motivations for adopting robotic mobile testing. 
Liu \etal \cite{jian2012application} discuss the application of tools in software test automation, with the Robot Framework automation test analysis. 
Cheng \etal \cite{cheng2021mobile} propose automated robot testing based on black-box testing combining the YOLOv3 algorithm and the perceptual hashing algorithm. 
Zhang \etal \cite{zhang2022machine} present MAEIT, which generate executable tests based on the mobile test robot by analyzing the video of the human-computer interactions and guiding the robot to complete the test task.

These applications of robotic arms in software testing show us the possibility of improving testing effectiveness beyond traditional intrusive testing frameworks. The limitations brought by the introduction of robotic arms motivate us to figure out an effective method to improve testing efficiency.

\subsection{Visual-based GUI Understanding}

Visual-based GUI understanding is a key part of our approach \cite{yu2019lirat}, and different visual analysis algorithms will have different degrees of impact on the results.
Emil \etal \cite{alegroth2015visual} propose that visual GUI testing is an emerging technology in industrial practice that offers greater flexibility and robustness to certain GUI changes than previous high-level GUI test automation techniques. 
Chen \etal \cite{chen2020object} introduce a new GUI-specific old-fashioned method for non-text GUI element detection using a novel top-down coarse-to-fine strategy, and incorporates it with the mature deep learning model for GUI text detection. 
Sato \etal \cite{sato2008support} propose a visualization technique that presents the correspondence between the screens before and after the operation, as well as the traces of the source code executed by the operation. 
Liu \etal \cite{liu2020discovering} propose OwlEye, a novel approach based on deep learning for modeling visual information of the GUI screenshot. 
Liu \etal \cite{liu2022nighthawk} further propose Nighthawk, a fully automated approach based on deep learning for modeling visual information of the GUI screenshot. 
Yu \etal \cite{yu2019crowdsourced} introduce BIU, which employs image understanding techniques to help crowdworkers automatically infer bugs and generate bug descriptions using app screenshots. 
Liu \etal \cite{liu2018generating} propose a fully automatic technique to generate descriptive words for well-defined screenshots by using the CV technique, namely spatial pyramid matching. 
Singh \etal \cite{singh2021surface} introduce a novel approach to achieve automation in the black-box environments by taking the app screenshots and understanding the widgets using image processing algorithms. 
Ren \etal \cite{ren2022cross} propose CdDiff to help testers improve mobile app compatibility by leveraging CV techniques. 
Borjesson \etal \cite{borjesson2012automated} present that visual GUI testing is applicable for automated system testing with effort gains over manual system test practices. 

The above studies open a door for utilizing GUI information purely from the visual perspective. Given the rapid advancement of CV technologies, both traditional algorithms and learning-based models, it is quite effective and practical to conduct the visual-based GUI understanding. Such a practice is really an appropriate solution to target widget recognition in non-intrusive GUI testing. However, all the studies focus on app screenshots, which are with high resolution. In this work, we focus on the camera-taken GUI images to make the CV technologies applicable in a more practical way for the robotic arm testing scenario.

\subsection{GUI Exploration Testing}


Azim \etal \cite{azim2013targeted} propose A3E, allowing Android apps to be explored systematically while running on actual phones. Two exploration strategies are developed to explore activities fast and directly. 
Salihu \etal \cite{salihu2016systematic} propose a hybrid approach for the systematic exploration of mobile apps, using static and dynamic approaches to improve the app's state exploration.
Model-based approaches \cite{amalfitano2012using, yang2013grey, wetzlmaier2016framework, yan2017widget, yan2018land, cao2018accelerating, lai2019goal} are deeply investigated for GUI exploration testing.
Model-based approaches are deeply investigated for GUI exploration testing.
Machiry \etal \cite{machiry2013dynodroid} propose a practical system Dynodroid for generating relevant inputs to mobile apps on the dominant Android platform.
Hao \etal \cite{hao2014puma} propose PUMA, a programmable UI automation framework for conducting dynamic analyses of mobile apps at scale.
Mao \etal \cite{mao2016sapienz} propose Sapienz, an approach to Android testing that uses multi-objective search-based testing to automatically explore and optimize test sequences. 
Su \etal \cite{su2017guided} propose Stoat, an approach to perform stochastic model-based testing on Android apps. It iteratively refines test generation to high coverage and diverse event sequences. 
Gu \etal \cite{gu2019practical} propose APE, a fully automated model-based tool for effective testing of Android apps. 
Lai \etal \cite{lai2019goal} propose GoalExplorer to generate executable test scripts that triggers the functionality as a solution for Android app exploration. 
Learning-based approaches \cite{zheng2021automatic, peng2022mubot, zhang2022unirltest} emerge due to the development of deep learning, machine learning, and reinforcement learning.
Zheng \etal propose Wuji \cite{zheng2019wuji}, a representative approach that uses reinforcement learning to explore video game apps. 
Q-testing \cite{pan2020reinforcement} is an effective method for Android testing with reinforcement learning, it trains a neural network to divide different states at the granularity of functional scenarios. 
Romdhana \etal \cite{romdhana2022deep} compare different RL algorithms in mobile app testing.

The above studies show the effectiveness on mobile app exploration testing in the intrusive scenario. However, in the non-intrusive, such approaches are weakened due to the extra time overhead brought by the robotic arm movement. This is the first work considering actual execution overhead with the PoP-guided exploration strategy to alleviate the efficiency challenge.

\section{Conclusion}

Faced with the challenges of GUI exploration testing, including the incapability to work with embedded industrial operating systems and the failure to identify compatibility bugs. \toolname is a practical non-intrusive solution that effectively, efficiently, and generally addresses these issues. \toolname employs a practical GUI exploration testing framework that utilizes visual-based robotic arms. It includes a set of advanced GUI screen and GUI widget identification technologies fitting the robotic arm testing scenario. Besides, a complete set of movements that flexibly simulate human testing interactions is designed for automated GUI exploration testing. In order to improve the testing efficiency of operations by robotic arms, \toolname involves a PoP-guided exploration strategy, which explores nearby widgets of previous operation targets to reduce the robotic movement overhead. For compatibility bug detection, we design a novel GUI comparison-based detection strategy. The experiment results show that \toolname effectively detects compatibility bugs beyond crash bugs. Besides, the PoP-guided exploration strategy helps improve testing efficiency.

\section*{Acknowledgment}
The authors would like to thank the anonymous reviewers for their insightful comments. This work is supported partially by the National Natural Science Foundation of China (62141215, 61932012, 62372228).


\bibliographystyle{ACM-Reference-Format}
\bibliography{main}

\end{document}